\documentclass[a4paper, runningheads]{llncs}

\newcommand{\orcidLM}	{\href{https://orcid.org/0000-0002-6866-0799}{\protect\includegraphics[scale=0.045]{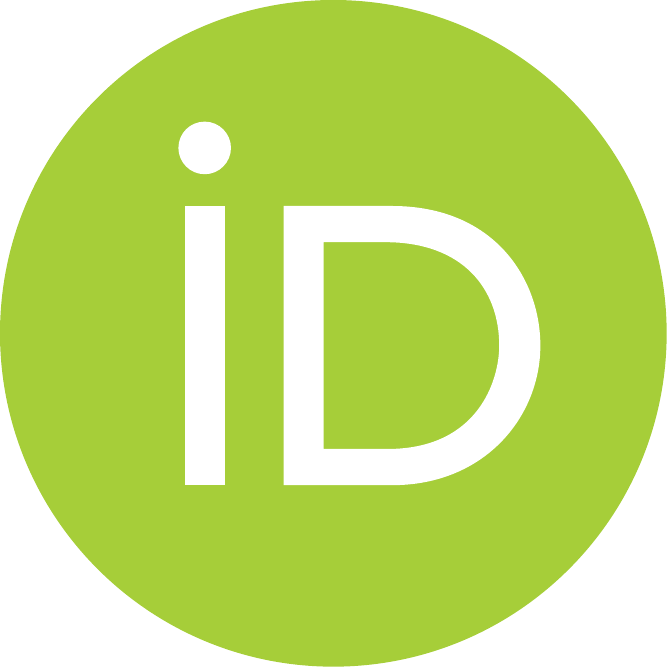}}}

\newcommand{\orcidJG}	{\href{https://orcid.org/0000-0002-7932-5822}{\protect\includegraphics[scale=0.045]{orcid.pdf}}}

\newcommand{\orcidJM}	{\href{https://orcid.org/0000-0002-6332-5801}{\protect\includegraphics[scale=0.045]{orcid.pdf}}}

\newcommand{\orcidSRM}	{\href{https://orcid.org/0000-0001-5656-6108}{\protect\includegraphics[scale=0.045]{orcid.pdf}}}

\newcommand{\orcidRB}	{\href{https://orcid.org/0000-0002-5515-7158}{\protect\includegraphics[scale=0.045]{orcid.pdf}}}

\newcommand{\orcidYB}   {\href{https://orcid.org/0000-0002-6407-7221}{\protect\includegraphics[scale=0.045]{orcid.pdf}}}

\newcommand{\orcidES}   {\href{https://orcid.org/0000-0001-7579-910X}{\protect\includegraphics[scale=0.045]{orcid.pdf}}}

\usepackage[USenglish]{babel}
\addto\captionsUSenglish{\renewcommand{\figurename}{Fig.}}
\usepackage{lmodern}
\usepackage{adjustbox}
\usepackage{tabto}
\usepackage{booktabs}
\usepackage{tabularx}
\usepackage{makecell}
\usepackage{rotating}
\usepackage{graphicx}
\usepackage{tikz}
\usepackage{pgfplots}
\pgfplotsset{compat=1.15}
\usepackage{multicol,multirow}
\usepackage{amssymb,amsmath}
\usepackage[autostyle, english=american]{csquotes}
\usepackage{url}
\usepackage{xcolor}
\usepackage{placeins}
\usepackage[inline, shortlabels]{enumitem}
\usepackage{cite}
\usepackage{breakcites}
\usepackage{listings}

\usepackage{todonotes}

\usepackage{xcolor}
\definecolor{lightgray}{rgb}{.9,.9,.9}
\definecolor{darkgray}{rgb}{.4,.4,.4}
\definecolor{forestGreen}{RGB}{34,139,34}
\definecolor{orangeRed}{RGB}{255,69,0}
\definecolor{codegreen}{rgb}{0,0.6,0}
\definecolor{codegray}{rgb}{0.5,0.5,0.5}
\definecolor{codepurple}{rgb}{0.58,0,0.82}

\usepackage{listings}   \lstdefinelanguage{xml}{
        basicstyle=\small,
        sensitive=false,
}

\lstdefinestyle{xmlStyle}{
escapechar=|,
language=xml,
frame=tb,
aboveskip=3mm,
belowskip=3mm,
basicstyle=\scriptsize,
sensitive=true,
showstringspaces=false,
numbers=left,
numberstyle=\tiny,
numberblanklines=false,
tabsize=4,
numbersep=3pt,
extendedchars=true,
xleftmargin=2em,
lineskip=1pt,
breaklines=true,
captionpos=b,
}
\lstnewenvironment{xml}[2]{
\lstset{caption=#1,label=#2,style=xmlStyle}
}{}

\RequirePackage[bookmarks,
                urlcolor=black,
                citecolor=black,
                linkcolor=black,
                pagecolor=black,
                colorlinks,
                hyperfigures]{hyperref}

\makeatletter
\def\RemoveSpaces#1{\zap@space#1 \@empty}
\makeatother

\makeatletter
\newcommand{\printfnsymbol}[1]{%
  \textsuperscript{\@fnsymbol{#1}}%
}
\makeatother

\newcommand{\eg}{e.\,g.,~}

\newcommand{\ie}{i.\,e.,~}

\begin{document}

\renewcommand{\thefigure}{\arabic{figure}}
\renewcommand{\thelstlisting}{\arabic{lstlisting}}

\title{SensorStream: An XES Extension for Enriching Event Logs with IoT-Sensor Data}

\titlerunning{SensorStream XES Extension}

\author{Joscha Grüger\thanks{\scriptsize{}Main authors in alphabetical order; All main authors contributed equally to this work.}\inst{1,2}\orcidJG \and
Lukas Malburg\printfnsymbol{1}\inst{1,2}\orcidLM \and
Jürgen Mangler\printfnsymbol{1}\inst{3}\orcidJM \and
Yannis Bertrand\inst{4}\orcidYB \and
Stefanie Rinderle-Ma\inst{3}\orcidSRM \and
Ralph Bergmann\inst{1,2}\orcidRB \and
Estefanía Serral Asensio\inst{4}\orcidES
}

\authorrunning{J. Grüger et al.}

\institute{Artificial Intelligence and Intelligent Information Systems, \\ University of Trier, 54296 Trier, Germany\\ \email{\{grueger,malburgl,bergmann\}@uni-trier.de}\\
\url{http://www.wi2.uni-trier.de} \\~%
\and German Research Center for Artificial Intelligence (DFKI) \\ Branch University of Trier, 54296 Trier, Germany\\ \texttt{\{firstname.lastname\}@dfki.de} \\~%
\and Department of Informatics, Technical University of Munich, \\85748 Garching, Germany
\\ \email{\{firstname.lastname\}@tum.de}\\~%
\and Research Centre for Information Systems Engineering (LIRIS), KU Leuven \\ Warmoesberg 26, 1000 Brussels, Belgium \\
\email{\{firstname.lastname\}@kuleuven.be}
}

\maketitle
\begin{abstract}
Process management and process orchestration/execution are currently hot topics; prevalent trends such as automation and Industry 4.0 require solutions which allow domain-experts to easily model and execute processes in various domains, including manufacturing and health-care. These domains, in turn, rely on a tight integration between hardware and software, \ie via the Internet of Things (IoT). While process execution is about actuation, \ie actively triggering actions and awaiting their completion, accompanying IoT sensors monitor humans and the environment. These sensors produce large amounts of procedural, discrete, and continuous data streams, that hold the key to understanding the quality of process subjects (\eg produced parts), outcome (\eg quantity and quality), and error causes. Processes constantly evolve in conjunction with their IoT environment. This requires joint storage of data generated by processes, with data generated by the IoT sensors is therefore needed. In this paper, we present an extension of the process log standard format XES, namely SensorStream. SensorStream enables to connect IoT data to process events, as well as a set of semantic annotations to describe the scenario and environment during data collection. This allows to preserve the full context required for data-analysis, so that logs can be analyzed even when scenarios or hardware artifacts are rapidly changing. Through additional semantic annotations, we envision the XES extension log format to be a solid based for the creation of a (semi-)automatic analysis pipeline, which can support domain experts by automatically providing data visualization, or even process insights.
\keywords{Process Management \and Industry 4.0 \and IoT Data \and Process Mining \and XES}
\end{abstract}
\section{Introduction}
\label{sec:Introduction}
In IoT environments where business processes are executed, a large amount of procedural data is generated. The use of these can enable the development of innovative applications in process control \cite{messner2019closed,Seiger.2022_IntegratingProcessManagement,Malburg.2020_FactoriesAndBPM}, process conformance checking \cite{ehrendorfer2019conformance,stertz2020data,stertz2020temporal}, or process enhancement \cite{pauker2021industry}, among others. Here, in particular, process mining techniques could be applied \cite{seiger2020IoTDrivenProcessEventLog,Seiger.2022_IntegratingProcessManagement,Malburg.2020_FactoriesAndBPM,stertz2020analyzing}. In contrast to many other domains, in the IoT domain, the context in which certain operations are performed is a particularly important factor \cite{Bertrand.2022_PMAndIoT}.

One domain where process orchestration/execution \cite{mangler2014cpee} and IoT meet is the manufacturing domain \cite{mangler2019centurio,pauker2021industry}. Here context is mainly derived from sensors that monitor the execution environment and resources during execution \cite{Sisinni.2018_IndustrialIoT,Seiger.2022_IntegratingProcessManagement,Elsaleh.2019_IoTStream,Janiesch.2020_Manifesto}.
In domains as manufacturing, process orchestration/execution \cite{mangler2014cpee,Seiger.2022_IntegratingProcessManagement,Malburg.2020_FactoriesAndBPM,mangler2019centurio,pauker2021industry} often relies on Internet of Things (IoT) technology. While IoT actuators can be used to automate process tasks, IoT sensors and tags can be used to closely monitor the execution environment and involved resources \cite{Sisinni.2018_IndustrialIoT,Seiger.2022_IntegratingProcessManagement,Elsaleh.2019_IoTStream,Janiesch.2020_Manifesto}. IoT technology can therefore capture the context in which certain process tasks are performed, which is a particularly important factor for techniques such as process mining \cite{Bertrand.2022_PMAndIoT,seiger2020IoTDrivenProcessEventLog} to better understand and analyze the manufacturing processes.
As such, besides the procedural data generated from the process execution, the data captured by IoT should also be considered an integral part of the process execution logs, i.e., the event logs.

Both the procedural nature of sensor logs and the tight integration of these with the execution processes and the executing resources \cite{ehrendorfer2021sensor}, makes sensor data an integral part of procedural application scenarios in IoT \cite{seiger2020IoTDrivenProcessEventLog,Seiger.2022_IntegratingProcessManagement,Bertrand.2022_PMAndIoT}. In this context,

From a process log perspective, IoT data can be assigned to different levels depending on its relation with the process and on the nature of the data collected (static vs dynamic, collection frequency, relation to the process, etc.). Depending on logging, knowledge about the executed processes and process models, and physical aspects such as the placement of sensors or their orientation, sensors can be directly assigned to individual events or traces or neither. In addition, IoT data is often ad hoc, highly variable, contains data quality issues, and has varying degrees of semantification \cite{Elsaleh.2019_IoTStream,BermudezEdo.2016_IoTLite}.

Thus, IoT event logs have special requirements on the data storage. Each observation of sensors must be assignable to events, traces, or neither, and  different degrees of semantification should be expressible with the inclusion of proper ontologies. However, the defacto standard for storing event logs in process mining, XES (eXtensible Event Stream)~\cite{gunther2014xes}, focuses in particular on the control flow perspective, not being able meet complex requirements on the data perspective \cite{Bertrand.2022_PMAndIoT}. In the absence of unified, expressive standards for IoT-enriched event logs, both industry and academia are developing their own proprietary formats. This results in many highly customized data formats and procedural applications for individual use cases that are not interoperable with each other.

In this paper, we present a new XES extension for uniform storage of IoT-enriched event logs. The extension complements XES in a way that extensive IoT sensor data can be stored on event, trace, or standalone. This extension is based on widely used ontologies, which enable the unified semantic enrichment of event logs based on a common vocabulary. The XES extension is intended to lay a foundation for process mining in IoT environments and to promote reusability and interoperability.

The structure of the paper is as follows: In Sect.~\ref{sec:Foundations}, we describe the theoretical basis for process mining in IoT and the related literature. Section~\ref{sec:XESExtensionForIoTEventLogs} introduces the proposed SensorStream XES extension to specify IoT-enriched event logs.  Section~\ref{sec:semantics} describes the annotation meta-model for using ontological information. In Sect.~\ref{sec:ApplicationScenario}, we present an application scenario for IoT-enriched event logs in smart manufacturing. Section~\ref{sec:ConclusionAndFutureWork} summarizes the results, lists advantages and limitations, and gives an outlook for future research directions.

\section{Foundations and Related Work}
\label{sec:Foundations}

The recent developments and technologies used in the \emph{Industrial Internet of Things (IIoT)} \cite{Sisinni.2018_IndustrialIoT} demand a more intelligent and interconnected process-based control of IoT devices \cite{Seiger.2022_IntegratingProcessManagement}. For a deeper integration with IoT environments in a process-oriented way, \emph{Business Process Management (BPM)} methods can be applied for control and analysis purposes \cite{Janiesch.2020_Manifesto}. The benefit of it is that BPM could profit from the huge variety of IoT sensor data that can be used to improve analysis methods. In the following three subsections, firstly  (Sect.\ref{subsec:PMforIoT}) we describe how process mining techniques can be applied to IoT environments, including the description of typical application scenarios. Secondly (Sect.~\ref{subsec:IoTOntologies}), we describe how ontologies are used in the IoT domain do improve the overall interoperability between systems and improve the understandability of IoT data. Thirdly (Sect. \ref{sec:Foundations:subsec:RelatedWork}), we describe related approaches that tackle data analysis problems, and provide data sets, which are related to the challenges and solutions described in this paper.

\subsection{Process Mining for IoT Environments}
\label{subsec:PMforIoT}
One way to analyze IoT sensor data and corresponding event log data is process mining. Process mining describes three analysis tasks. The most common is (i) Process Discovery. Discovery techniques take an event log and produce a process model from it \cite{vanderAalst.2011,BERTI.2019}. The second task is (ii) Conformance Checking, which is used to validate the conformance of real process instances to a given a-priori model. The last analysis task is (iii) Enhancement that uses an event log and the associated process model to identify bottlenecks and, thus, to improve the process accordingly \cite{vanderAalst.2011}. Tasks i-iii are based on an event log as input. An event log can contain one or more traces. Each trace represents a process instance. A trace, in turn, consists of the sequence of executed activities, each represented by an event. Furthermore, event logs can store additional attributes, such as timestamps, resources, and data elements \cite{vanderAalst.2018b}. Many proposals have been made in the past for storing process logs. MXML as a simple XML format for audit and trails in process ware information systems \cite{MXML.2005}. XES, the current standard event log model, is also based on XML and widely used in both industrial and academic contexts \cite{Dongen2015RelationalXD,gunther2014xes}.

The XES metamodel can be represented in XML. An XES attribute consist of (a) a data type represented by the qualified name of the XML element, (b) a key to denote the type of attribute (unique within its container), and (c) a value (see Listing \ref{lst:AttributeExample}). XES describes six types of attributes: string, date, int, float, boolean and id which have a value, as well as two additional attributes, container and list, which can hold arbitrary child attributes. All attributes can also be nested (even inside non-container and non-list attributes) \cite{gunther2014xes}.

\begin{lstlisting}[caption={Sample XES (XML serialization) with Trace, Events and Attributes.},label=lst:AttributeExample]
<log xes.version="1.0"
     xmlns="http://www.xes-standard.org"
     xes.creator="cpee.org"
     xes.features="nested-attributes">
  <extension name="Concept" prefix="concept" uri="http://www.xes-standard.org/concept.xesext"/>
  <extension name="Lifecycle" prefix="lifecycle" uri="http://www.xes-standard.org/lifecycle.xesext"/>
  <extension name="Identity" prefix="identifier" uri="http://www.xes-standard.org/identity.xesext"/>
  <extension name="Time" prefix="time" uri="http://www.xes-standard.org/time.xesext"/>
  <global scope="trace">
    <string key="concept:name" value="name"/>
  </global>
  <global scope="event">
    <string key="concept:name" value="name"/>
    <string key="lifecycle:transition" value="start"/>
    <date key="time:timestamp" value="1970-01-01T00:00:00.000+00:00"/>
  </global>
  <string key="lifecycle:model" value="standard"/>
  <string key="creator" value="cpee.org"/>
  <string key="library" value="cpee.org"/>
  <trace>
    <string key="concept:name" value="Process 1"/>
    <event>
      <string key="concept:name" value="Task 1"/>
      <string key="lifecycle:transition" value="start"/>
      <date key="time:timestamp" value="1970-01-01T00:00:00.000+00:00"/>
      <string key="name" value="Juergen"/>
    </event>
    <event>
      <string key="concept:name" value="Task 2"/>
      <string key="lifecycle:transition" value="start"/>
      <date key="time:timestamp" value="1970-01-01T00:00:00.000+00:00"/>
      <string key="name" value="Juergen"/>
    /event>
  </trace>
</log>
\end{lstlisting}

Since the requirements for event logs differ depending on the application and domain, XES can be extended via so-called extensions. Standard extensions include the concept extension, which specifies a generally understood name for events, traces, or the log. In addition, the lifecycle extension can be used to specify different stages in the lifecycle of events and the time extension standardizes the specification of event timestamps \cite{gunther2014xes}.
XES also allows the definition of new data attribute types through the notion of extensions, thereby increasing the flexibility of the model.

Several implementations of the standard coexist, the main one being OpenXES \footnote{\url{https://www.xes-standard.org/openxes/start}}, which is used by many event logs described in the literature.

Recently, the uptake of new technologies and the gain in maturity of the process mining field have increased the urge to create alternative event log models. Multiple propositions that relax some assumptions of XES and allow for more flexibility in event data storage have been presented (\eg \cite{Popova2013,OCELStandard2021}). Among them, a standard for \emph{Object-Centric Event Logs (OCEL)} \cite{OCELStandard2021} has been developed to be more suitable for storing event data extracted from relational databases and is widely considered as the main challenger of XES today. OCEL replaces the strict notion of case with the concept of object, which generalizes it by allowing one event to be linked with multiple objects instead of a single case. This removes the necessity to "flatten" the event log by picking one case notion from the several potential case notions that often coexist in real-life processes. A second noticeable difference with XES is the explicit inclusion of the concept of activity in OCEL, which is absent in XES.

\subsection{Ontologies for IoT-Environments}
\label{subsec:IoTOntologies}
To represent knowledge about the IoT environment and also about the data that is produced in it, several ontologies have been proposed.
The focus in these IoT ontologies has shifted from trying to be as complete as possible (\eg the \emph{Semantic Sensor Network (SSN)} ontology \cite{ComptonBBGCCGHHHHJKPLLNNPPST12SSNontology} or the \emph{CREMA Data Model, Core module (CDM-Core)} \cite{DBLP:conf/ic3k/MazzolaKVK16CDM-Core}) to be simpler and more practical in real world applications (\eg the IoT-Lite \cite{BermudezEdo.2016_IoTLite} ontology). This also follows the industry need for analytics and real-time processing techniques, among which we can find process mining techniques. One such ontology is the \emph{Sensors, Observations, Samples, and Actuators (SOSA)} ontology \cite{Janowicz2019}. SOSA is a compact version of the SSN ontology and describes the relationships between sensors and actuators as well as their measured observations in IoT data. By using this ontology, it is possible to represent, for example, a relationship between a machine resource and the sensors that monitor its condition. A further ontology especially tailored for streaming data is the IoT-Stream ontology \cite{Elsaleh.2019_IoTStream}. It is a more specific ontology, inspired by SOSA, that focuses on the treatment of streaming data. Both the SOSA and the IoT-Stream ontologies are event-centric, in the sense that they focus on data generation and treatment, and less attention is paid to the devices and platforms on which IoT relies, such as in the CDM-Core ontology.

\subsection{Related Work}
\label{sec:Foundations:subsec:RelatedWork}

Recently multi-perspective process mining \cite{mannhardt_multi-perspective_2018}
has evolved which, for example, uses process data as presented in \cite{stertz_detecting_2019}. Also, the
analysis of time series data is used as described in
\cite{stertz_detecting_2019} and \cite{SMR2020} for detecting concept drifts
during run-time. A survey on outcome-oriented predictive process monitoring
presented in \cite{10.1145/3301300} compares different techniques.

For all of these approaches, data sets have been provided containing a wealth of context data in conjunction with process events. These datasets however present slightly different granularity levels, slightly different formats, and slightly different semantics.

To the best of our knowledge, no fine-grained semantics enabled format to unify IoT data and process data storage has been proposed.

\section{An XES-Extension for IoT-Enriched Event Logs}
\label{sec:XESExtensionForIoTEventLogs}

XES is built around events, which describe how a sequence of activities has
been executed. Each activity can be lead to a set of events in a XES log file, following
the life-cycle (see Sect. \ref{sec:Foundations}) of the execution of that activity in a particular instance. I.e., each activity could lead to a \textit{``start''} event, to a \textit{``complete''} event, and to an arbitrary number of events in between, depending on the utilized life-cycle model.

Many XES log files just store one event per executed activity, thus sensor readings can be attached to this event. Other available logs, such as \cite{stertz2020temporal}, expose a custom fine-grained life-cycle model, that anchors sensor reading to an event with special XES lifecycle:transition.

\begin{figure}[ht]
  \centering
  \includegraphics[width=0.5\linewidth]{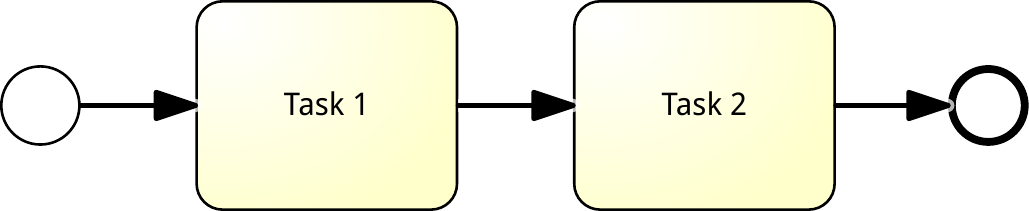}
  \caption{Example Process}
  \label{fig:example1}
\end{figure}

The case shown in Fig.~\ref{fig:example1} leads to the XES
log described in Fig.~\ref{lst:AttributeExample}. As mentioned in the XES Standard:

\begin{quote}
  \enquote{Log, trace, and event objects contain no information themselves.
  They only define the structure of the document. All information in an event
  log is stored in attributes. Attributes describe their parent element (log,
  trace, etc.). All attributes have a string-based key.}
\end{quote}

We propose an extension to the XES standard in order to capture and store IoT
based sensor data within the XES log file for future enhanced analysis, data
mining, and process enhancement purposes (cf. process mining; see Sect.~\ref{subsec:PMforIoT}).

We see the XES file as a long-term storage format that can hold all aggregated
data connected to a process. When assuming heterogeneous sources for all
connected data artifacts, it is also fair to assume that some of these sources
will evolve and change their structure. Thus, holding the extracted, transformed
and aggregated data in a flexible, structure long-term storage format is
imperative.

\begin{figure}[ht]
  \centering
  \includegraphics[width=0.7\linewidth]{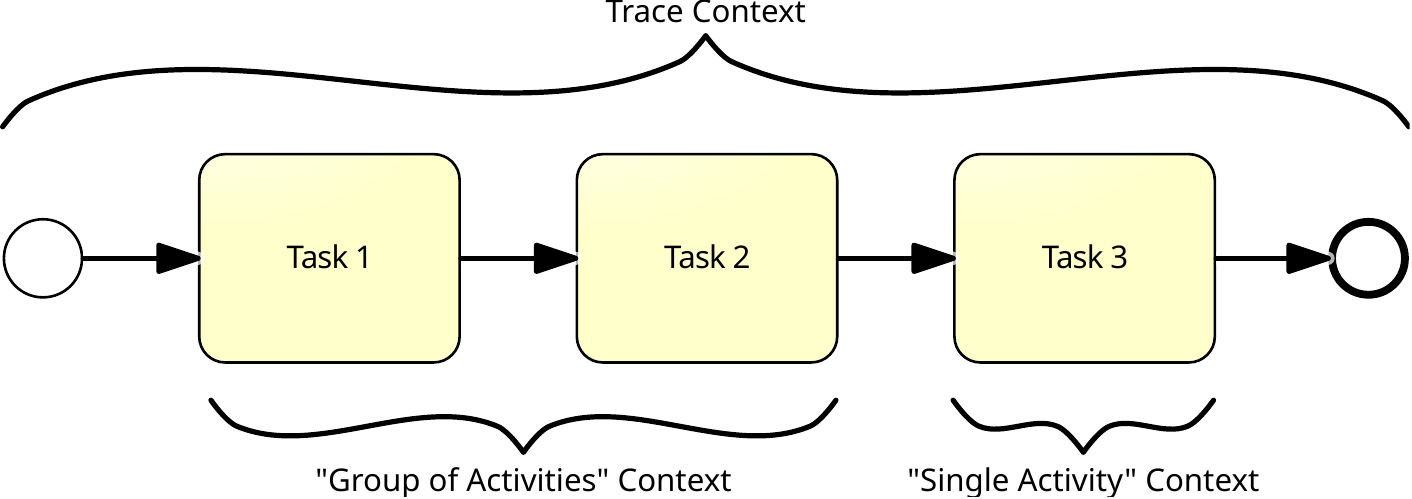}
  \caption{Different Contexts in Which IoT Data Can Be Collected}
  \label{fig:example2}
\end{figure}

In order to understand how IoT data might be connected to process activities,
we have to distinguish between three different cases (see
Fig.~\ref{fig:example2}):

\begin{itemize}

    \item \textbf{"Single Activity" Context}: A time-series of sensor readings
    from at least one sensor is connected to a single activity, \eg when the
    activity represents the machining of a part, collected sensor data might
    describe various aspects, such as the throughput of coolant while
    machining, a discrete series of vibration readings, or a function
    (continuous data) describing the noise generation (volume). All sensor data
    can be assigned to a particular activity, only data between the start and
    the completion of the activity is relevant.

    \item \textbf{"Group of Activities" Context}: A time-series of sensor
    readings from at least one sensor is connected to a set of activities. This
    is especially relevant for environmental sensors (which again might provide
    discrete or continuous readings), which for example span a multitude of
    production steps. These steps occur in parallel, thus leading to a variety
    of possible traces but the temperature development might give insights into
    certain quality properties of a finished product.

    \item \textbf{"Trace" Context}: A time-series of sensor readings from at least one
    sensor is connected to a whole trace. This case is analogous to the "Group
    of Activities" case. Assignment to individual activities is not possible or
    not desired, as \eg the sensor readings in order to provide proper analysis
    context may have to contain a period before and a period after individual
    activities.

\end{itemize}

\begin{figure}[ht]
  \centering
  \includegraphics[width=\linewidth]{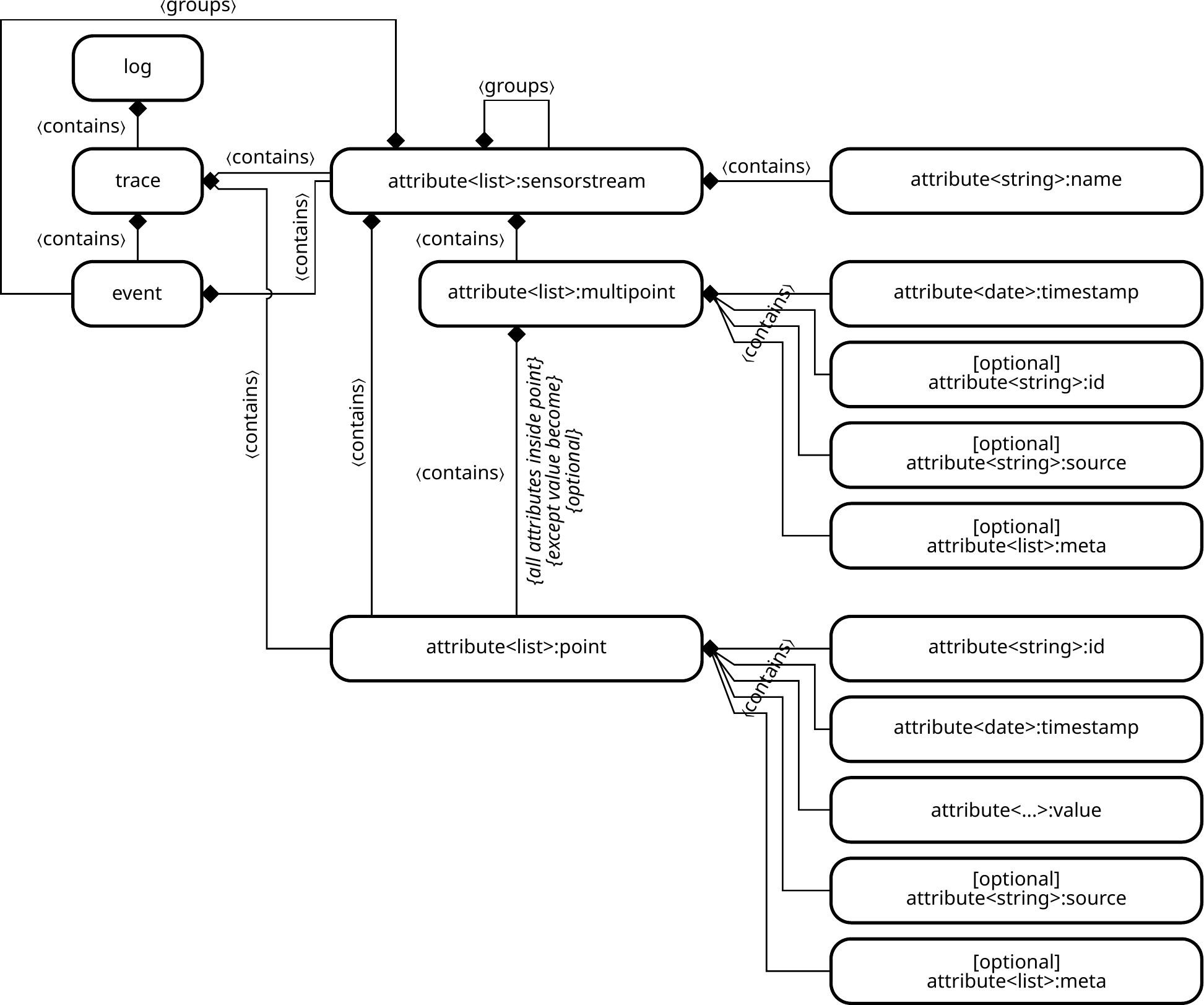}
  \caption{SensorStream Metamodel}
  \label{fig:meta}
\end{figure}

In order to realize these three contexts, and to store individual sensor readings, we propose the following meta-model depicted in Fig.~\ref{fig:meta}.

In the following, we will denote all attributes of our proposed extension with the prefix \textbf{stream:}, to facility the clarity of the description. We will furthermore assume that the \textbf{stream:} prefix is specified in a XES extension \url{https://cpee.org/sensorstream/sensorstream.xesext}.

The core of the extension is stream:point. It contains all the attributes
that allow us to represent individual sensor values as XES artifacts. It is a
list. Values include:

\begin{itemize}

    \item \textbf{id}: uniquely identifies the sensor, \eg if a gyro-sensor
    delivers orientation and angular velocity changes separately, the
    identifiers can be gyro/velocity and gyro/angular\_velocity. On the other
    hand, if the sensor delivers a value pair, the identifier can be gyro.

    \item \textbf{source}: identify the source of a sensor value, \eg a
    drilling machine is the source of many different sensor readings at all
    times. The source attribute allows grouping these values into groups that
    might belong together and, thus, make sense to be analysed together. The
    source is optional.

    \item \textbf{timestamp}: A timestamp when the reading was taken. The
    timestamp is intended to be in ISO 8601 format, including milliseconds
    (YYYY-MM-DDTHH:mm:ss.sssZ) or microseconds (YYYY-MM-DDTHH:mm:ss.ssssssZ).

    \item \textbf{value}: The value delivered by the sensor. As sensors can
    deliver single values (float, int, strings) or complex data (pairs,
    triplets, deeply structured data, \ldots), we always assume this is stored
    as some serialized string representation. How to support the automatic
    extraction of potential parts of the value as well as semantic meaning will
    be discussed in Sect.~\ref{sec:semantics}.

    \item \textbf{meta}: A straightforward extension point, which allows to
    specify an additional list of attributes, which might be important for
    custom data analysis purposes. Meta is optional.

\end{itemize}

\subsection{Context, Grouping, and Nesting: \textit{stream:point}, \textit{stream:sensorstream}}

The second introduced concept (see Fig.~\ref{fig:meta}) is the
stream:sensorstream. It was introduced as the missing piece to produce
\textbf{"Single Activity"}, \textbf{"Group of Activities"}, and \textbf{"Trace"} contexts. Its only (optional) attribute is name, which can be used to describe the purpose of the grouping.

If a set of stream:points is included directly in the level of the trace, all
points are meant to exist in the \textbf{"Trace"} context: they cannot be
attributed to any event or group of events yet.

If a stream:sensorstream exists at the trace level, the stream:sensorstream has to group multiple
events, and it has to contain at least one stream:point. This realizes the
\textbf{"Group of Activities"} context. Multiple stream:sensorstream attributes
can exist at trace level, meaning that multiple groups exists.

If a stream:sensorstream exists at the event level, it has to contain at least
on stream:point. Multiple stream:sensorstream can exist at the event level.
While this does not change the meaning of all these points being connected to
one event, its purpose might be to further structure the events, \eg separating
two different levels of importance for analysis purposes.

All stream:sensorstream attributes might be nested. While this is not very
interesting, at the event level, it is important at the trace level. Nested
sensor:sensorstream attributes might convey different layers of connection
granularity. For example, some stream:point attributes might be grouped to a
group [1] of 2 tasks, some other stream:point attributes might be connected to
a group [2] of 2 different tasks. Then a third set of stream:point attributes
might be connected to all tasks in groups [1] and [2], leading to a [3: [1]
[2]] nesting, as depicted in List. \ref{lst:nest}:

\begin{lstlisting}[caption={Sample XES (XML serialization) stream:sensorstream Nesting},label=lst:nest]
  <trace>
    <string key="concept:name" value="Process 1"/>
    <list key="stream:sensorstream">
      <list key="stream:point">
        <date key="stream:timestamp" value="2021-11-04T15:22:19.367+01:00"/>
        <string key="stream:id" value="humidity"/>
        <string key="stream:value" value="62.5"/>
      </list>
      [...]
      <list key="stream:sensorstream">
        <list key="stream:point">
          <date key="stream:timestamp" value="2021-11-04T15:22:22.369+01:00"/>
          <string key="stream:id" value="pressure"/>
          <int string key="stream:value" value="19"/>
        </list>
        <event>[...]</event>
        <event>[...]</event>
        [...]
      </list>
      <list key="stream:sensorstream">
        <list key="stream:point">
          <date key="stream:timestamp" value="2021-11-04T15:22:28.369+01:00"/>
          <string key="stream:id" value="temperature"/>
          <int string key="stream:value" value="75.3"/>
        </list>
        <event>[...]</event>
        <event>[...]</event>
        [...]
      </list>
    </list>
  </trace>
</log>
\end{lstlisting}

This leaves us with the special case of overlapping cases, where some
stream:point's are connected to tasks 1 and 2, where some other stream:point's
are connected to tasks 2 and 3. This case can only (XES being a tree structure)
be solved by creating three stream:sensorstream attributes with some duplicated
stream:point elements.

\subsection{Convenience and Storage Size: \textit{stream:multipoint}}

The final element introduced in Fig.~\ref{fig:meta} is stream:multipoint. This
concept is not necessary from a functional perspective, but allows reducing the
size of the log file.

For example, when a set of sensor:point attributes all origin from the same
sensor and the same source, and contain the same meta information, this
information is duplicated all over and over. A sensor:multipoint allows to
group this redundant information for a set of points:

\begin{lstlisting}[caption={Sample XES (XML serialization) stream:multipoint},label=lst:multipoint1]
  <trace>
    <string key="concept:name" value="Process 1"/>
    <event>
      <string key="concept:name" value="Task 1"/>
      <string key="lifecycle:transition" value="transition"/>
      <date key="time:timestamp" value="1970-01-01T00:00:00.000+00:00"/>
      <string key="name" value="Juergen"/>
      <list key="stream:sensorstream">
        <string key="stream:name" value="Temperature"/>
        <list key="stream:multipoint">
          <string key="stream:id" value="keyence/mesurement"/>
          <string key="stream:source" value="keyence"/>
          <list key="stream:point">
            <date key="stream:timestamp" value="2021-11-04T15:22:19.367+01:00"/>
            <string key="stream:value" value="18"/>
          </list>
          <list key="stream:point">
            <date key="stream:timestamp" value="2021-11-04T15:22:20.369+01:00"/>
            <int string key="stream:value" value="19"/>
          </list>
        </list>
      </list>
    </event>
  </trace>
</log>
\end{lstlisting}

Alternatively, it can be used to group according to timestamp, if a set of
sensor readings are taken at discrete points in time:

\begin{lstlisting}[caption={Sample XES (XML serialization) stream:multipoint},label=lst:multipoint2]
  <trace>
    <string key="concept:name" value="Process 1"/>
    <event>
      <string key="concept:name" value="Task 1"/>
      <string key="lifecycle:transition" value="transition"/>
      <date key="time:timestamp" value="1970-01-01T00:00:00.000+00:00"/>
      <string key="name" value="Juergen"/>
      <list key="stream:sensorstream">
        <string key="stream:name" value="Temperature"/>
        <list key="stream:multipoint">
          <date key="stream:timestamp" value="2021-11-04T15:22:19.367+01:00"/>
          <list key="stream:point">
            <string key="stream:id" value="temperature"/>
            <string key="stream:value" value="48.5371"/>
          </list>
          <list key="stream:point">
            <string key="stream:id" value="pressure"/>
            <string key="stream:value" value="12:30-1,12:31-2,3,4,5"/>
          </list>
        </list>
      </list>
    </event>
  </trace>
</log>
\end{lstlisting}

\section{Facilitating Analysis: Conveying Sensor Data Semantics}
\label{sec:semantics}

Based on the XES extension presented in
Sect.~\ref{sec:XESExtensionForIoTEventLogs}, it is possible to properly store
and attribute/connect/correlate sensor data to process logs. While this is a
proper basis for data analysis (conformance and compliance checking, root cause
analysis in case of bad quality of process subjects, \ldots), various analysis
tasks still require further preparation:

\begin{itemize}

  \item Structural understanding of sensor:point values, their units/dimensions.

  \item Semantic understanding of the scenario, in which the data has been collected.

  \item Semantic understanding of the meaning of sensor:point values, why and
  how they are connected to the scenario or parts of it.

  \item Semantic relation between different sensors.

  \item Semantic understanding of how sensor:point values, or their changes, relate to different sensor:point values and their changes.

\end{itemize}

Such analysis tasks probably require domain knowledge. As our
goal is to facilitate analysis tasks as much as possible, it is important to
also store information about semantic aspects in the log.
This will allow stored data to be self-contained objects for analysis, so that even after long-term storage as much semantic knowledge as possible can be preserved.

\begin{figure}[ht]
  \centering
  \includegraphics[width=1.0\linewidth]{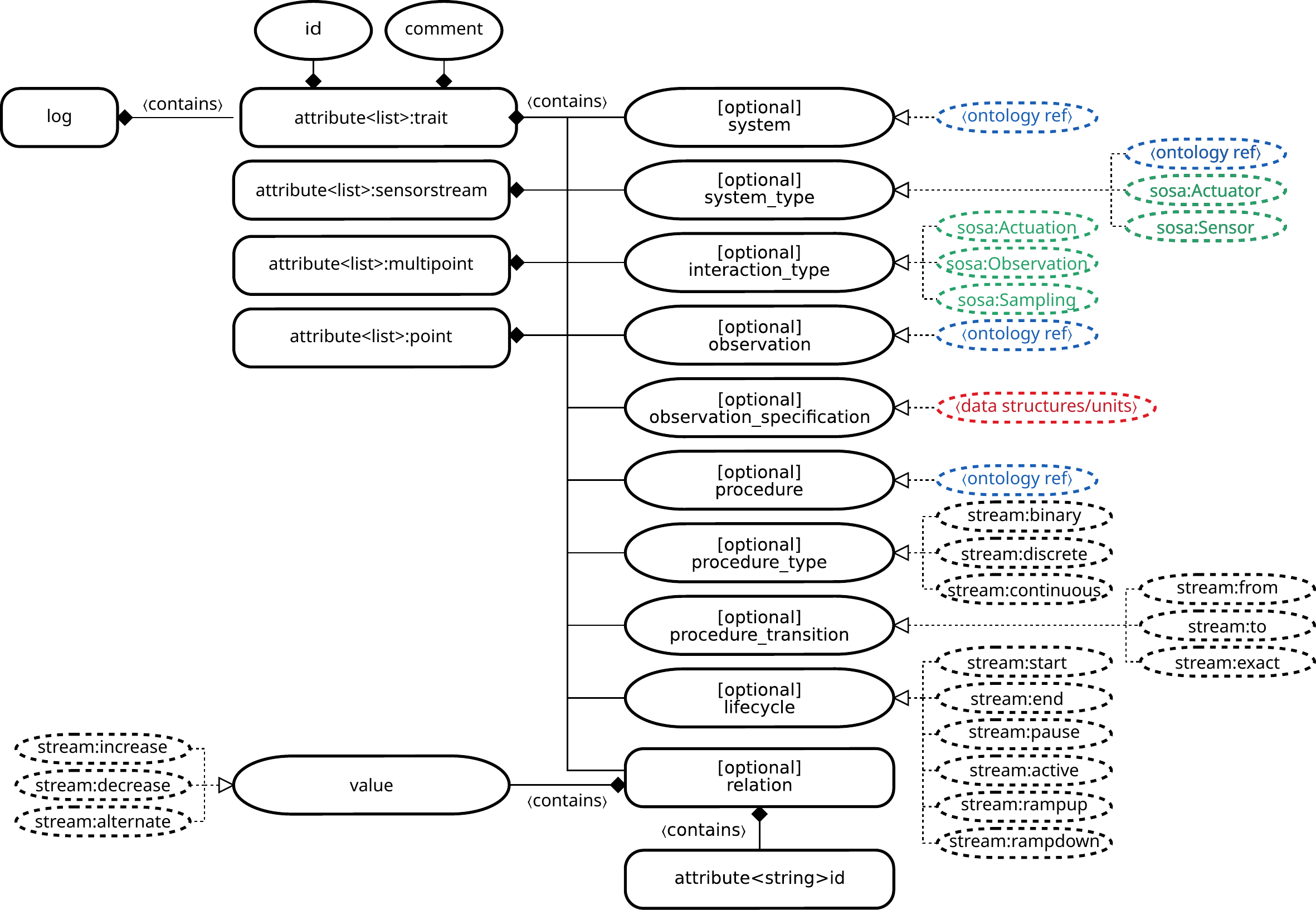}
  \caption{SensorStream Annotation Meta-Model}
  \label{fig:annometa}
\end{figure}

In order to do so, we developed an annotation meta-model that can be used to
connect ontological information to stored XES logs, as depicted in Fig.
\ref{fig:annometa}.

As depicted in List. \ref{lst:annometa}, the extension is based on existing IoT
ontology standards (SOSA and SSN; see Sect.~\ref{subsec:IoTOntologies}), and
realized as a set of annotation based on XML namespaces.

\begin{lstlisting}[caption={Sample XES (XML serialization) stream:multipoint},label=lst:annometa]
<log xes.version="1.0"
     xmlns="http://www.xes-standard.org"
     xes.creator="cpee.org">
     xmlns:stream" uri="https://cpee.org/sensorstream/"
     xmlns:ssn" uri="http://www.w3.org/ns/ssn/"
     xmlns:sosa" uri="http://www.w3.org/ns/sosa/">
  <extension name="SensorStream" prefix="stream" uri="https://cpee.org/sensorstream/sensorstream.xesext"/>
  <extension name="Concept" prefix="concept" uri="http://www.xes-standard.org/concept.xesext"/>
  <extension name="Lifecycle" prefix="lifecycle" uri="http://www.xes-standard.org/lifecycle.xesext"/>
  <extension name="Identity" prefix="identifier" uri="http://www.xes-standard.org/identity.xesext"/>
  <extension name="Time" prefix="time" uri="http://www.xes-standard.org/time.xesext"/>
</log>
\end{lstlisting}

\subsection{Semantic Annotation of stream:point}

In order to convey additional semantic information, stream:point (and analogously
stream:multipoint, stream:sensorstream) can be annotated, with a wide variety
of information describing the many semantic aspects of the data collection.
Please note that assigning the information to stream:multipoint or
stream:sensor\-stream just means that the information is true for all points.

List.~\ref{lst:seman} shows a sample annotation to a single point by using the
developed domain ontology FTOnto~\cite{Klein.2019b_FTOnto} as a concrete
implementation of a specific knowledge representation:

\begin{lstlisting}[caption={Sample XES (XML serialization) With Semantic Annotations},label=lst:seman]
<list key="stream:point"
      stream:system="http://iot.uni-trier.de/FTOnto#OV_1"
      stream:system_type="http://iot.uni-trier.de/FTOnto#Oven"
      stontologyream:system_type="sosa:Actuator"
      stream:interaction_type="sosa:Observation"
      stream:observation="http://iot.uni-trier.de/FTOnto#Temperature"
      stream:observation_specification="[temperature,temperature] [degree celsius,degree celsius]"
      stream:procedure="http://iot.uni-trier.de/temperature_measurement_method"
      stream:procedure_type="stream:discrete"
      stream:procedure_transition="stream:exact"
      stream:lifecycle value="stream:active">
  <string key="stream:id" value="temperature/>
  <stream:relation value="stream:increase">
    <string key="stream:id" value="pressure/>
  </stream:relation>
  [...]
</list>
\end{lstlisting}

As can be seen in List.~\ref{lst:annometa}, and List. \ref{lst:seman}, we rely
heavily on parts of the \emph{Sensors, Observations, Samples, and Actuators
(SOSA)} ontology \cite{Janowicz2019} and the \emph{Semantic Sensor Network
(SSN)} ontology \cite{ComptonBBGCCGHHHHJKPLLNNPPST12SSNontology}, which
themselves are closely related.

We especially utilize the concepts described by these standards to clarify the
involved systems (SSN term), nature of observations, and procedures.

All annotations are optional and should only be added when known and
unambiguous. The following annotations were inspired either by SOSA or SSN:

\begin{itemize}

  \item \textbf{system}: Link to an ontology that describes the type of system
  the value stems for. In the example above, the FTOnto domain ontology
  describes an oven.

  \item \textbf{system}: Type of system that delivers the value. This can be a
  sosa:actuator which yields a value on actuation, a sosa:sensor.
  Alternatively, a link to an ontology can be provided, which can in turn
  describe a sosa:sensor or a sosa:actuator.

  \item \textbf{interaction\_type}: Type of interaction that happened in order
  to read the value. A sosa:actuation can very well return a value describing
  its state, while sosa:observation describes a single sensor reading.
  sosa:sampling describes a set of subsequent sosa:observations.

  \item \textbf{observation}: Link to an ontology describing the nature of the
  value, \eg a temperature reading.

  \item \textbf{observation\_specification}: Intended to provide a simple
  machine-readable description of the structure of the value returned by a
  sensor. This covers individual values, lists of values and their units.  For
  anything more complex, the information should be provided in
  \textbf{observation}. In the example above, the syntax becomes clear:
  [temperature] describes that a single value is delivered,
  [temperature,temperature] describes that a pair of values is delivered, and
  so forth. The second expression, separated by a comma, assigns units (like
  \eg m/s) to each denoted value.

  \item \textbf{procedure}: A link to an ontology describing how exactly the measurement has been taken, \eg by describing hardware properties, timing, \ldots:

\end{itemize}

All other annotations do not have equivalent concepts in SSN or SOSA, and allow
interpreting the nature of values, \eg to automatically select the right
visualization for a set of sensor:point attributes:

\begin{itemize}

  \item \textbf{procedure\_type}: Allows to annotate the type of
  data/measurement. Data might be stream:binary (\eg on/off), stream:discrete
  if the measurements are taken periodically by the sensor\footnote{This might
  include internal storage; when the sensor is queried, it delivers the last
  stored reading; alternatively the sensor might push the value immediately
  after the reading is taken.}, or stream:continuous. While a value has always
  a timestamp attached, stream:continuous points at analogue sensors, which
  might yield functions describing the progression since the last measurement,
  or data aggregations. Thus, when interpreting this stream:continuous data, it
  is possible that outliers are missing, or have been factored into the data.

  \item \textbf{procedure\_transition}: Points at how the progression between
  different values of this sensor is affecting the underlying business process. A domain expert might point out patterns that show data changes (e.g. a simple value switching from off to on) which are clearly connected to tasks in a process. For example, a task triggering the start of motor might always lead to a set of very specific data changes, which clearly point out the existence of the task, even if the actual actuation (or the logging of the task in a process engine) can not be observed. Patters can include: (1) stream:from points at progression in the process when the current value
  changes. (2) stream:to points to a progression when the current value changes to   a different value, set of values or range. (3) stream:exact finally describes a  progression exactly from the current value, to a different value, set of  values or range.

  \item \textbf{lifecycle}: Describes in which phase of the process a sensor value occurs, \eg when an oven is at a temperature between 80 and 100
  degrees Celsius that means it is active. Again, a domain expert might easily point out a set/range of values that can allow to partition data and deduce the existence of tasks, even if they can not be directly observed.  Consequently, stream:start denotes a
  value which might indicate that something has just been started (works also
  for stream:binary). stream:end on the other hand describes a value that
  denotes the end of a task, (or process). stream:pause denotes a set of values
  that points towards a break in underlying process or production.
  stream:active points at the sensor measuring activity, \eg a vibration sensor
  might measure transport. stream:rampup points to values that denote, \eg a
  motor spinning up or an oven pre-warming, while stream:rampdown points to the
  opposite.

\end{itemize}

All of the above values might allow to mark data points or sets of data-points
in a visualization with their respective meaning. The annotation of course
requires, that when the data is written, a component which holds
domain-specific knowledge, interprets data values and adds the appropriate
annotations.

Through the presented annotations, the formalization of such domain specific
knowledge can be structured and simplified. By interpreting the annotations,
universal analysis components can automatically provide improved
visualizations.

\subsection{Interaction: stream:relation}

In order to provide a means to describe the relation between stream:point
attributes, inside of stream:point it is possible to insert a stream:relation
attribute. The stream:relation attribute describes how a particular point can
be interpreted in relation to another stream:point from a different sensor.
stream:increase denotes that when this value increases, the value of the other
sensor should also increase, \eg temperature and pressure. stream:decrease
describes the opposite. stream:alternating describes a relationship where
increases and decreases should alternate.

Through this attribute, simple consistency and conformance checking becomes
possible.

\subsection{Reusability: stream:trait}

The stream:trait attribute can be used at log level to create a template that
can be reused for all stream:point attributes. A stream:trait acts a anchor point for attaching all possible semantic annotations, which then in turn can be referenced by arbitrary stream:point attributes via a custom stream:trait XML attribute, as depicted in List.~\ref{lst:trait}.

This again decreases the size of the log file and thus improves
parsing speed.

A sample trait integration is depicted in List.~\ref{lst:trait}:

\begin{lstlisting}[caption={Sample XES (XML serialization) stream:trait},label=lst:trait]
<log xes.version="1.0"
     xmlns="http://www.xes-standard.org"
     xes.creator="cpee.org">
     xmlns:stream" uri="https://cpee.org/sensorstream/"
     xmlns:ssn" uri="http://www.w3.org/ns/ssn/"
     xmlns:sosa" uri="http://www.w3.org/ns/sosa/">
  [...]
  <stream:trait id="1" comment="protocol properties"
      stream:procedure_type="stream:binary"
      stream:procedure_transition="stream:exact"
  </stream:trait>
  <stream:trait id="1" comment="other properties"
    [...]
  </stream:trait>
  [...]
  <trace>
    <list key="stream:point" stream:trait="1 2">
      [...]
    </list>
  </trace>
</log>
\end{lstlisting}

At the level of a stream:point XES attribute multiple traits can be reference
by a stream:trait XML attribute.

\section{Application Scenario for IoT-Enriched Event Logs in Smart Manufacturing}
\label{sec:ApplicationScenario}
In order to evaluate our proposed SensorStream XES extension, we will present and discuss an IoT-enriched event log from a physical smart factory in this section. First, we present how the smart factory for event log generation is structured and controlled and how the log has been generated (see Sect.~\ref{subsec:PhysicalFactorySimulationModels}). Afterwards, we show a part of the enriched event log and describe use cases for process mining analysis (see Sect.~\ref{subsec:log}).

\subsection{Physical Factory Simulation Lab Utilizing BPM Technology at the University of Trier}

\label{subsec:PhysicalFactorySimulationModels}
\begin{figure}[!htb]
  \centering
  \includegraphics[width=0.9\textwidth]{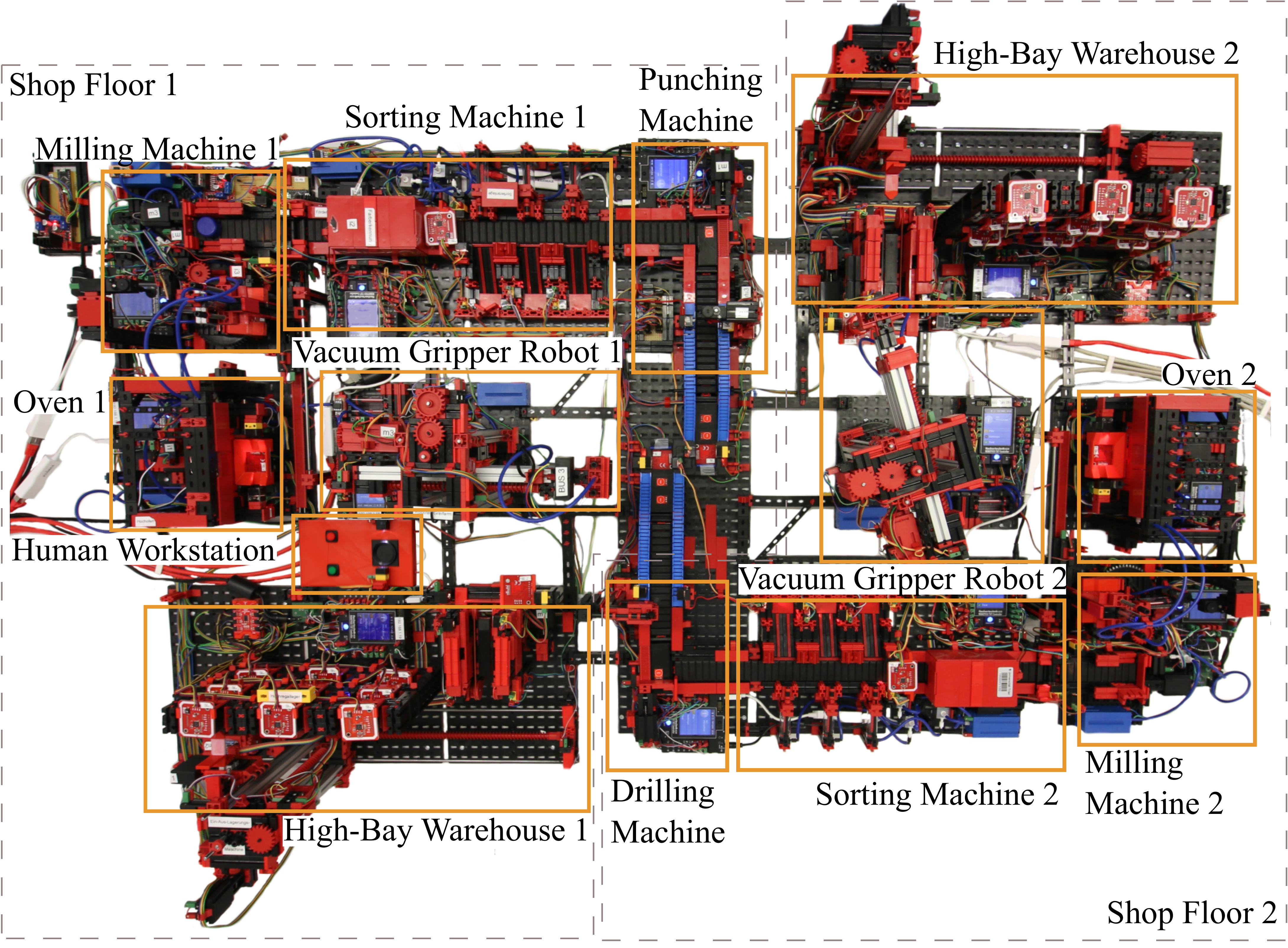}
  \caption{The Physical Factory Simulation Model. (Source: \cite{Malburg.2020_FactoriesAndBPM})}
  \label{img:grafik-factoryModel}
\end{figure}

\emph{Learning Factories} \cite{Abele.2017_LearningFactories} are gaining importance in Industry~4.0 research (cf.~\cite{seiger2020IoTDrivenProcessEventLog,Malburg.2020_FactoriesAndBPM,Seiger.2022_IntegratingProcessManagement}). In contrast to widely used artificially generated data, physical simulation models provide much more realistic data and run-time behavior such as ad-hoc interventions \cite{Malburg.2020_FactoriesAndBPM}. Thus, they enable the transfer to real-world settings more easily. In our research \cite{Klein.2019_PredM,Malburg.2020_FactoriesAndBPM,Malburg.2020_SemanticWebServices_IN4PL,Seiger.2022_IntegratingProcessManagement,Malburg.2021_ObjectDetection,Hoffmann.2022_ProGAN}, we use a physical factory simulation model from Fischertechnik. The custom model\footnote{\url{https://iot.uni-trier.de}} simulates two independently working production lines consisting of two shop floors that are linked for the exchange of workpieces. Figure~\ref{img:grafik-factoryModel} illustrates the used factory simulation model.
By using a BPM abstraction layer \cite{Seiger.2022_IntegratingProcessManagement,Malburg.2020_FactoriesAndBPM,Malburg.2020_SemanticWebServices_IN4PL}, it is possible to control the factory model in a process-based fashion by using workflow management systems. A video from the factory executing a manufacturing process can be found in \cite{Malburg.2020_ObjectDetectionDemoVideo}. During process execution, the generated data (event log and IoT sensor data) is stored using the proposed SensorStream XES extension format. In the following, we describe use cases for process mining analysis based on the generated data from process executions.

\subsection{IoT-Enriched Event Log}
\label{subsec:log}
The event log contains X traces and N events, which in turn are assigned Z sensor observations. Since the event log was built with background knowledge about the processes and thus top-down, there is only sensor data at event level. List.~\ref{lst:enreve} shows the structure of a single event in an event log. Using the concept, time and organizational extension, the activity, start time, and the resource \textit{vacuum gripper robot 1} are given.
\begin{lstlisting}[caption=Structure of a Trace and Event,label={lst:enreve}]
<trace>
        <event>
            <string key="concept:name" value="calibrating motor 3"/>
            <string key="org:resource" value="vgr_1"/>
            <date key="time:timestamp" value="2021-06-25T17:08:50.414000"/>
            <date key="operation_end_time" value="2021-06-25T17:08:57.163000"/>
            <list key="stream:sensorstream">
                [...]
           </list>
            [...]
\end{lstlisting}
In List.~\ref{lst:sensorstr_accel} a section of the sensor data of the previously shown activity is shown. For the description and exact mapping of the acceleration sensor, the stream system refers to the FTOnto ontology \cite{Klein.2019b_FTOnto}. More generally, the sensor is semantically classified by the system\_type. The observation indicates via a corresponding link to the ontology that an acceleration has been observed. The observation is further specified in the observation\_specification, which indicates that the observation consists of the points x, y, and z, which are measured in $m/s^2$. Procedure\_type indicates that they are continuous values and interaction\_type says that it is an observation. Finally, time and measured values are given. Observing all three given measurements, the movement of the vacuum gripper robot can be traced on the X, Y, and Z axes.

\begin{lstlisting}[caption={Excerpt of the Acceleration Sensors’ Data, Related to the Motor Calibration Shwon in List. \ref{lst:enreve}},label={lst:sensorstr_accel}]
<list key="stream:sensorstream">
    <list key="stream:point" stream:system="http://iot.uni-trier.de/FTOnto#BMX055_Pi_1_AccSensor_1" stream:system_type="sosa:Sensor" stream:observation="http://iot.uni-trier.de/FTOnto#VGR_1_Crane_Jib_Acceleration" stream:observation_specification="[x,y,z] [m/s^2,m/s^2,m/s^2]" stream:procedure_type="stream:continuous" stream:interaction_type="sosa:Observation">
        <date stream:timestamp="2021-06-25T17:08:50.414718"/>
        <string stream:value="[-0.721, -10.2483, -0.7114]"/>
    </list>
    <list key="stream:point" stream:system="http://iot.uni-trier.de/FTOnto#BMX055_Pi_1_AccSensor_1" stream:system_type="sosa:Sensor" stream:observation="http://iot.uni-trier.de/FTOnto#VGR_1_Crane_Jib_Acceleration" stream:observation_specification="[x,y,z] [m/s^2,m/s^2,m/s^2]" stream:procedure_type="stream:continuous" stream:interaction_type="sosa:Observation">
        <date stream:timestamp="2021-06-25T17:08:50.417484"/>
        <string stream:value="[0.4711, -9.46, -1.0575]"/>
    </list>
    <list key="stream:point" stream:system="http://iot.uni-trier.de/FTOnto#BMX055_Pi_1_AccSensor_1" stream:system_type="sosa:Sensor" stream:observation="http://iot.uni-trier.de/FTOnto#VGR_1_Crane_Jib_Acceleration" stream:observation_specification="[x,y,z] [m/s^2,m/s^2,m/s^2]" stream:procedure_type="stream:continuous" stream:interaction_type="sosa:Observation">
        <date stream:timestamp="2021-06-25T17:08:50.419621"/>
        <string stream:value="[-0.7691, -10.1426, -0.25]"/>
    </list>
    [...]
    \end{lstlisting}

Listing~\ref{lst:sensorstrpneu} shows data from the motor's speed sensor on vacuum gripper robot 1. Unlike the acceleration sensor data, only one sensor value is observed here. This is shown in the value and observation\_specification. Moreover, the value comes directly from the resource, without being a directly addressable or writable sensor. Therefore, the SOSA ontology is used here to explicitly describe that it is an actuation.
\begin{lstlisting}[caption={Excerpt of the Acceleration Sensors’ Data, Related to the Motor Calibration Shwon in List. \ref{lst:enreve}},label={lst:sensorstrpneu}]
<list key="stream:point" stream:system="http://iot.uni-trier.de/FTOnto#VGR_1_Motor_3" stream:system_type="sosa:Actuator" stream:observation="http://iot.uni-trier.de/FTOnto#MotorSpeed" stream:procedure_type="stream:continuous" stream:interaction_type="sosa:Actuation">
    <date stream:timestamp="2021-06-25T17:09:10.383000"/>
    <string stream:value="0.0"/>
</list>
\end{lstlisting}

\FloatBarrier
\subsection{Results}
\label{subsec:Results}
The use of the SensorStream extension enabled the integrated of all sensor data with the available process execution events. The extensive semantic description of the data via linkable ontologies proved to be well applicable. For individual sensors that are not fully semantically modeled in the domain ontology of the smart factory (see Sect.~\ref{subsec:PhysicalFactorySimulationModels}), the flexibility of the extension to include free text or references to entities in other ontologies (\eg SOSA) proved to be useful.

Limitations exist so far in the usability of the data structure in analysis tools, as there are no implementations to use it. Moreover, due to the large amounts of data in the IoT context, the logs can quickly become very large, therefore, scalability mechanisms should be further investigated.

\FloatBarrier
\section{Conclusion and Future Work}
\label{sec:ConclusionAndFutureWork}
In this paper, an extension to XES has been presented, allowing the joint storage of process event logs and IoT data related to the environment where these events occur.

The extension identifies what is required from the IoT perspective to enable the use of BPM methods for IoT (cf. \cite{Janiesch.2020_Manifesto}). As it has been shown,  there are special requirements for the data perspective in the IoT context, especially with respect to sensor data.

In the future, a complete event log of a factory shall be parsed and visualized based on the proposed extension format, to support process refinement and root cause analysis.

\iffalse
\begin{itemize}
    \item The extension should show what is needed from the IoT perspective to enable the usage of BPM methods for IoT
    \item This extension could be a first step for the current efforts to improve the current XES standard
    \item FW: Propose an event log using the extension (factory log)
    \item FW: Integrate with the event abstraction approaches
    \item FW: More radically new standard?
    \item FW: Visualization
    \item FW: Process Mining methods for IoT-enriched event logs
\end{itemize}
\fi

\interlinepenalty10000
\bibliographystyle{splncs04}
\bibliography{bibliography.bib,tum.bib}

\end{document}